\documentclass[prl,twocolumn,reprint,graphicx,showpacs,superscriptaddress,floatfix]{revtex4-1}
\usepackage{epsfig}
\usepackage{amsmath}
\usepackage{amssymb}
\usepackage{amsfonts}
\usepackage{mathptmx}
\usepackage{dcolumn}
\usepackage{eucal}
\usepackage{bm}
\usepackage{color}
\usepackage[colorlinks,linkcolor=blue,citecolor=blue]{hyperref}

\usepackage{epstopdf}

\usepackage{graphicx}

\usepackage{natbib}

\newcommand{\tmpnote}[1]%
   {\begingroup{\color{blue}\it (FIXME: #1)}\endgroup}

\newcommand{\mathE}{\mathcal{E} }

\begin{document}
\title{Metallic  supercurrent field-effect transistor}

\author{Giorgio De Simoni}
\affiliation{NEST, Instituto Nanoscienze-CNR and Scuola Normale Superiore, I-56127 Pisa, Italy}
\author{Federico Paolucci}
\affiliation{NEST, Instituto Nanoscienze-CNR and Scuola Normale Superiore, I-56127 Pisa, Italy}
\author{Paolo Solinas}
\affiliation{SPIN-CNR, Via Dodecaneso 33, 16146 Genova, Italy}
\author{Elia Strambini}
\affiliation{NEST, Instituto Nanoscienze-CNR and Scuola Normale Superiore, I-56127 Pisa, Italy}
\author{Francesco Giazotto}
\email{francesco.giazotto@sns.it}
\affiliation{NEST, Instituto Nanoscienze-CNR and Scuola Normale Superiore, I-56127 Pisa, Italy}

\maketitle

\textbf{In their original formulation of superconductivity, the London brothers predicted \cite{London1935} the exponential suppression of an \emph{electrostatic} field inside a superconductor over the so-called London penetration depth, $\lambda_L$ \cite{Hirsch1,Hirsch2,Tinkham}.
Despite a few experiments indicating hints of  perturbation induced by electrostatic fields \cite{Tao1,Glover1960,NbCluster}, no clue has been provided so far on the possibility to manipulate metallic superconductors via field-effect.
Here we report field-effect control of the supercurrent in \emph{all}-metallic transistors made of different Bardeen-Cooper-Schrieffer (BCS) superconducting thin films. 
At low temperature, our field-effect transistors (FETs) show a monotonic decay of the critical current under 
increasing electrostatic field up to total quenching for gate voltage values as large as $\pm 40$V in titanium-based devices. 
This \emph{bipolar} field effect 
persists up to $\sim 85\%$  of the critical temperature ($\sim 0.41$K), and in the presence of sizable magnetic fields. 
A similar behavior was observed in aluminum thin film FETs.
A phenomenological theory 
accounts for our observations, and points towards the interpretation in terms of 
an electric-field-induced perturbation propagating inside the superconducting film. In our understanding, this affects the pairing potential and  quenches the supercurrent. 
 These results could represent a groundbreaking asset for the realization of an \emph{all}-metallic superconducting  field-effect electronics and leading-edge quantum information architectures \cite{Marcus1,Marcus2}.
}

%%%%%%%%%%%%%%%%%%%%%%%%%%%%%%%%%%%%%%%%%%%%%%%%%%%%%%%%%%%%%%%%%%%
\begin{figure}[t!]
\begin{center}
\includegraphics[width=\columnwidth]{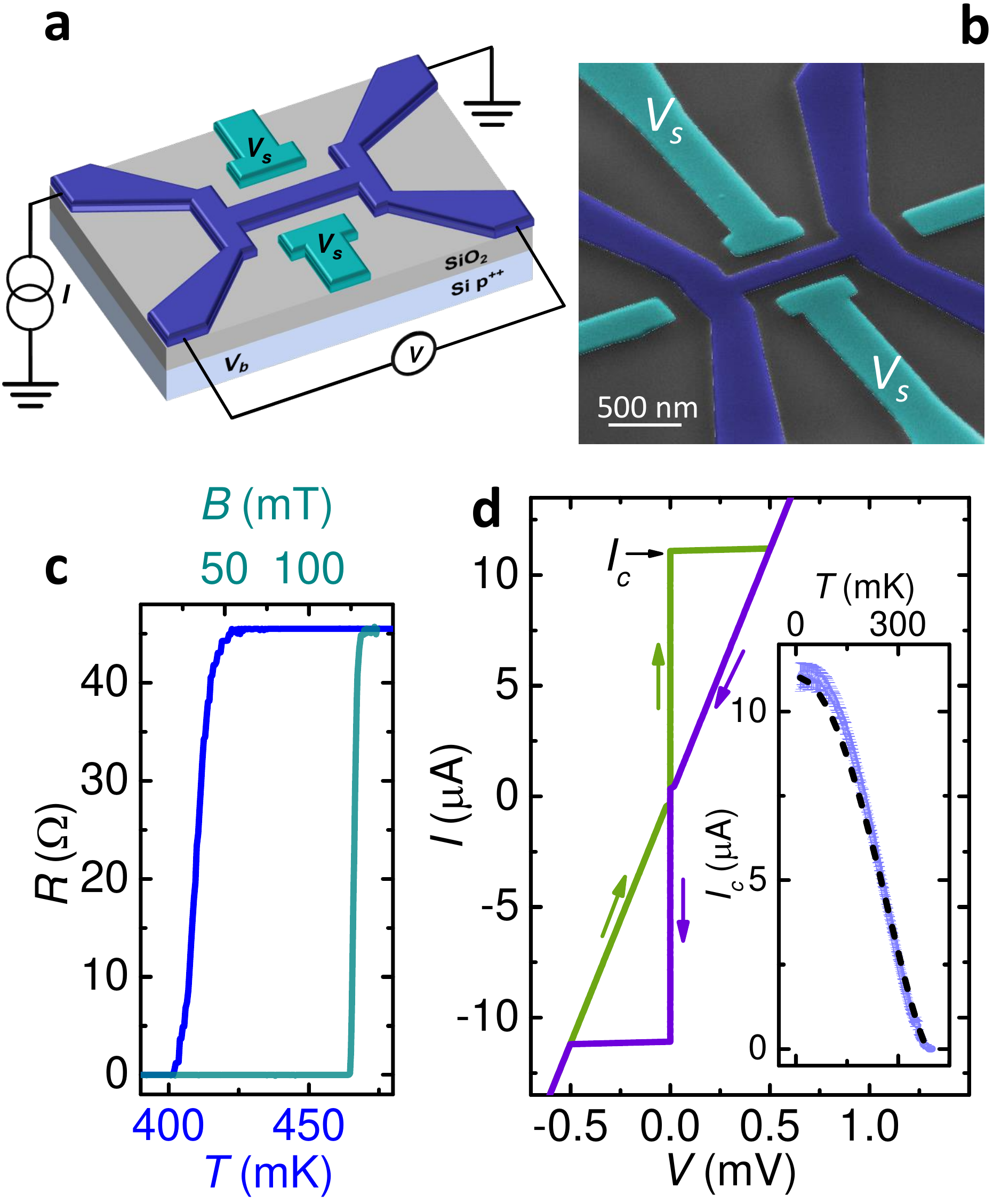}\vspace{-3mm}
\caption{\textbf{The metallic supercurrent field-effect transistor: pre-characterization.} 
\textbf{a}, Schematic of the \emph{all}-metallic supercurrent FET. 
Back and side gate voltage denoted with $V_b$ and $V_s$, respectively, are used to control the amplitude of the Ti wire critical supercurrent. $V_b$ is applied to the $p^{++}$ Si substrate.
In four-probe measurements, an electric current $I$ is fed into the superconducting wire whereas the voltage drop $V$ is simultaneously recorded as a function of the applied gate voltage. 
\textbf{b}, Pseudo-color scanning electron micrograph of a representative Ti supercurrent FET. 
The wire length is $900\,$nm while the width is $200\,$nm, and the thickness is $30\,$nm. 
The transistor core is shown in blue whereas the Ti side gates are colored in cyan. 
\textbf{c}, Resistance $R$ vs temperature $T$ (blue line, bottom horizontal axis), and $R$ vs perpendicular-to-plane magnetic field $B$ at $5$mK (light blu line, top horizontal axis) characteristics of the supercurrent FET. The normal-state resistance of the device is $R_N\sim 45\Omega$.
\textbf{d}, FET $I$ vs $V$ characteristic measured at $5\,$mK of bath temperature. 
$I_c\simeq 11\,\mu$A denotes the wire switching critical current, corresponding to $\sim 1.8 \times 10^5$Acm$^{-2}$ current density consistent with Ti in this temperature range \cite{Bardeen1962}.
The inset shows the full temperature evolution of $I_c$, and its quenching at the critical temperature ($T_c \sim 410$mK). Dashed line is the prediction of Ref. \cite{Bardeen1962}.   
The error bars represent the standard deviation of the critical current  $I_c$  calculated over 50 measurements.
}
\label{fig1}
\end{center}
\end{figure}
%%%%%%%%%%%%%%%%%%%%%%%%%
%%%%%%%%%%%%%%%%%%%%%%%%%%%%%%%%%%%%%%%%%%
%%%%%%%%%%%%%%%%%%%%%%%%%%%%%%%%%%%%%%%%%%%%%%%%%%%%%%%%%%%%%%%%%%% 
%%%%%%%%%%%%%%%%%%%%%%%%%%%%%%%%%%%%%%%%%%%%%%%%%%%%%%%%%%%%%%%%%%%
\begin{figure*}[ht!]
\begin{center}
\includegraphics[width=0.85\textwidth]{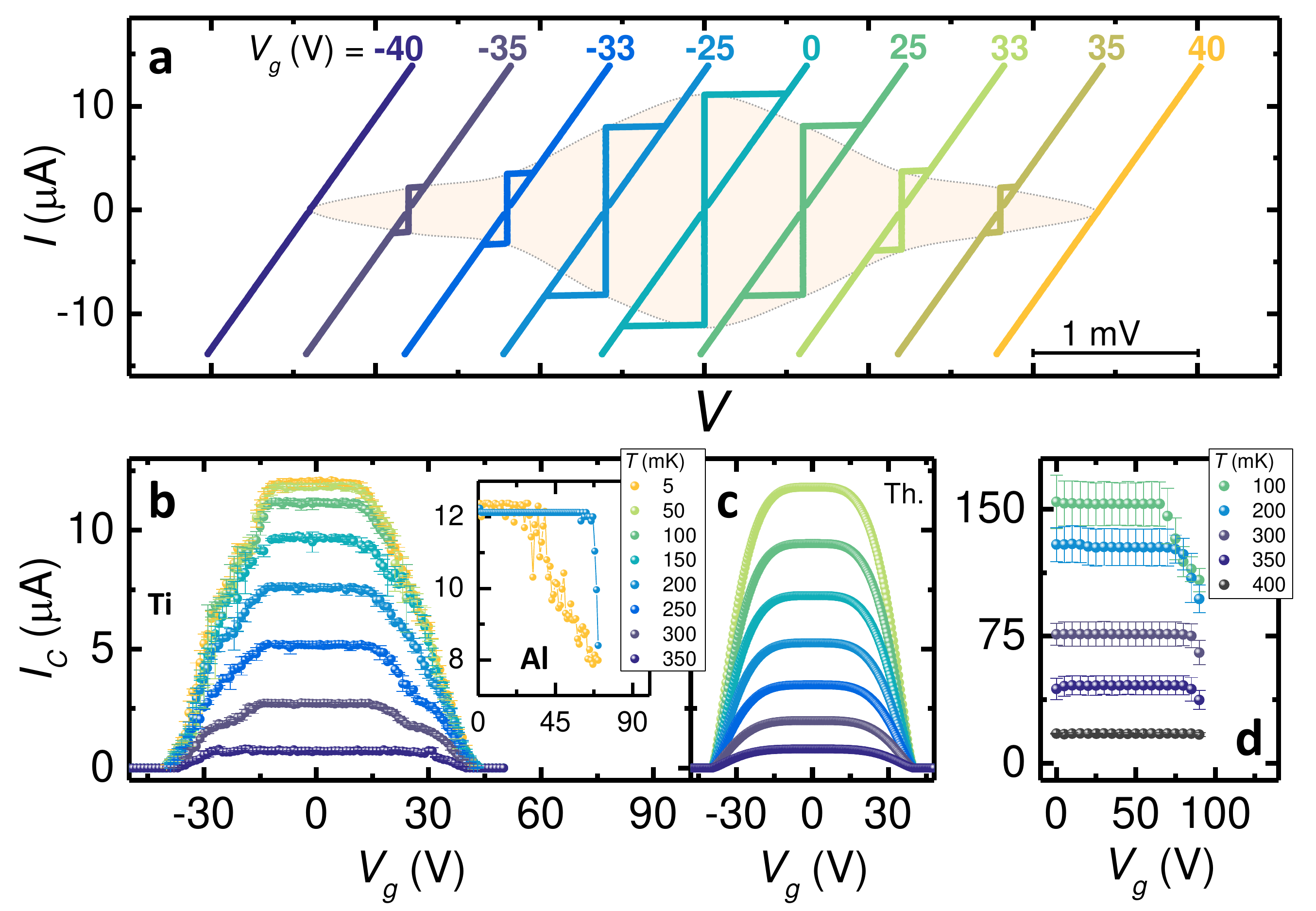}\vspace{-3mm}
\caption{\textbf{Electrostatic-field dependence of the supercurrent FET.} 
\textbf{a}, Current-voltage ($I-V$) characteristics of a Ti supercurrent FET measured at $5\,$mK for several values of the applied gate voltage ($V_g\equiv V_b=V_s$). 
Hatched area is a guide to the eye emphasizing the monotonic suppression of the critical current $I_c$ down to zero by increasing $|V_g|$. 
When the supercurrent vanishes, the $I(V)$ characteristics  coincide with that in the normal state, independently of the value of the applied gate voltage.
The curves are horizontally offset for clarity. 
\textbf{b}, Behavior of the critical current $I_c$ vs $V_g$  measured at different bath temperatures $T$. 
Note the full suppression of $I_c$ occurring for $|V_g|\gtrsim 40\,$V. This threshold voltage turns out to be almost independent of bath temperature.
The inset shows the $I_c(V_g)$ characteristics for a supercurrent transistor made of aluminum (Al) at two different bath temperatures. Here, the switching current is as large as $\simeq 12.3 \mu$A, corresponding to $\simeq 3.7 \times 10^6$Acm$^{-2}$ current density in Al \cite{Bardeen1962,Anthore2003}.
Note that the different shape of the $I_c(V_g)$ curves in Ti and Al might be related to a different temperature evolution of the electrostatic screening in the two superconducting materials.
\textbf{c}, Phenomenological critical current $I_c$ vs $V_g$ characteristics calculated with an \emph{ad hoc} model based on  Ginzburg-Landau formalism for the same values of bath temperature as in panel b. 
We emphasize  the agreement with the experimental data provided by the model, both for the monotonic decay of the $I_c(V_g)$  characteristics and for the temperature independence of the threshold voltage ($V_g^c$) over which full suppression of $I_c$ occurs. 
\textbf{d}, $I_c$ vs $V_g$ characteristics  measured at different bath temperatures $T$ of a $4\mu$m$\times 4\mu$m Ti FET showing  electrostatic field-tuning of the critical current occurring in a large (with respect to $\lambda_L$ and $\xi_0$) superconducting structure. 
In this case, the maximum achieved $I_c$ relative reduction  is around $\sim 30\%$ at $100$mK.
Here, the electrostatic field was applied only through the back gate ($V_g\equiv V_b$).
A critical temperature $T_c\simeq 470$mK, and a perpendicular-to-plane magnetic critical field $B_c\simeq 115$mT are the main characteristic parameters of the Ti film realizing this FET.
The error bars in panel b and d represent the standard deviation of the critical current $I_c$ calculated over 50 and $20$ measurements, respectively.
}
\label{fig2}
\end{center}
\end{figure*}

A generic scheme of our all-metallic supercurrent FETs is displayed in Fig. \ref{fig1}a. The transistors have been fabricated by electron-beam lithography and evaporation of titanium (Ti) or aluminum (Al) on top of an \emph{p}$^{++}$-doped Si substrate covered by a 300-nm-thick SiO$_2$  insulating layer (see Methods Summary for further details). 
Back and side gate voltages, denoted  by $V_b$ and $V_s$, respectively, are used to generate the electrostatic field that
controls the supercurrent flow in the wire.
Figure \ref{fig1}b shows a pseudo-color scanning electron micrograph of a typical device. 
The Ti FET consists of a wire of length $l=900$nm, width $w=200$nm, and thickness $t=30$nm. Similar transistors with length up to a $3\mu$m have shown  the same behavior. 
From the critical temperature $T_c\simeq 410$ mK  [see Fig. \ref{fig1}c showing the wire resistance ($R$) vs temperature ($T$) characteristic (blue line)] we determined the zero-temperature BCS energy gap, $\Delta_0=1.764 k_B T_c\simeq 62 \mu$eV, where $k_B$ is the Boltzmann constant.  
With this parameter, and from the low-temperature normal-state resistance of the wire, $R_N\simeq 45\Omega$, we deduced the London penetration depth, $\lambda_L=\sqrt{\hbar R_N wt/(\pi \mu_0 l \Delta_0)}\simeq 900$nm where $\mu_0$ is the vacuum magnetic permeability, and the superconducting coherence length $\xi_0=\sqrt{\hbar l/(R_NwtN_Fe^2\Delta_0)}\simeq 100$nm where $N_F\simeq 1.35\times 10^{47}$J$^{-1}$m$^{-3}$ is the density of states at the Fermi level of Ti and $e$ is the electron charge.
Since $w,t\ll  \lambda_L$,
the superconductor can be uniformly penetrated by external magnetic fields \cite{Anthore2003}.   
Figure \ref{fig1}c displays the wire  $R$ vs perpendicular-to-plane $B$ characteristic  at $5$mK (light blue line) revealing a  magnetic critical field as large as $\sim 127$mT.

Below $T_c$, dissipationless charge transport occurs in the transistor through Cooper pairs supercurrent. 
The wire current-voltage ($I-V$) characteristics recorded at $5$mK is shown in the main panel of Fig. \ref{fig1}d. 
In particular, a switching critical current with amplitude $I_c\simeq 11\mu$A is observed displaying an hysteretic behavior which might stem from heating induced in the wire while switching from the resistive to the 
dissipationless regime \cite{Courtois2008}. 
The monotonic decay of $I_c$ as a function of temperature, and its full  quenching at $T_c$ is shown in the inset of  Fig. \ref{fig1}d along with the phenomenological prediction by Bardeen (dashed line), $I_c(T)=I_c^0[1-(T/T_c)^2]^{3/2}$ where $I_c^0$ is the zero-temperature, zero-gate voltage wire critical current \cite{Bardeen1962}.

Investigation of field effect in our system is performed by current biasing the superconducting wire, and by measuring the  critical current vs gate voltage  ($V_{g}$) applied simultaneously to the back and side gates so that $V_b=V_s\equiv V_{g}$ (see Fig. \ref{fig1}a). 
This gate bias configuration maximized the impact of the electric field on $I_c$. 
Figure \ref{fig2}a shows the transistor $I-V$ characteristics measured at $5$mK for increasing values of $|V_{g}|$. 
The critical current is monotonically reduced by increasing $V_{g}$, and is fully suppressed at a critical voltage $|V_{g}^c|\simeq 40$V. 
At a first glance, this $I_c$ behavior might resemble  the one achieved in semiconductor-based Josephson FETs (JoFETs) \cite{Marcus1,Marcus2,Takayanagi1985,Kleinsasser1989,Doh2005,Xiang2006} though our transistors are fully \emph{metallic} so that the electric field does not affect the $I-V$ characteristic when the wire switches into the normal state.
Yet, in contrast to semiconductor-based JoFETs, the  effect is also \emph{bipolar}, i.e., symmetric in  the polarity of $V_{g}$. 
The experimental evidence of $I_c$ reduction and quenching seems to point towards the interpretation in terms of a field-effect-induced deformation and suppression of the superconducting pairing potential in the wire, and is the hallmark of the impact of a static electric field on a metallic superconducting  film.

The full temperature dependence of field effect in the Ti transistor is shown in Fig. \ref{fig2}b that displays the $I_c$ vs $V_{g}$ characteristics for several bath temperatures. 
By increasing $T$ the critical current plateau widens whereas,  at the same time, the critical voltage $V_g^c$ determining full suppression of  $I_c$ is almost temperature independent. 
Notably, field effect persists up to about $\sim 85\%$ of $T_c$, i.e., $\sim 350$mK, 
and proves the impact of the electric field on $I_c$ occurring even at higher temperatures.

Analogous behavior was also observed in supercurrent  FETs made of thin film aluminum (Al), and consisting of a wire of length $l=800$nm, width $w=30$nm, and thickness $t=11$nm. 
Key parameters of these structures are a normal-state resistance $R_N\simeq 320\Omega$, a critical temperature $T_C\simeq 1.5$K,
a London penetration depth $\lambda_L\simeq 310$nm, and a coherence length $\xi_0\simeq 63$nm (see SI). 
The Al FETs were realized  without side gates so that only a back gate $V_g\equiv V_b$ is used to control $I_c$.
The inset of Fig. \ref{fig2}b displays the $I_c$ vs $V_g$  characteristics at two selected bath temperatures.  Specifically, a monotonic suppression of $I_c$ from $\simeq 12.3\mu$A at $V_g=0$ down to $\simeq 8\mu$A for $V_g=70$V (corresponding to a relative suppression of the order of $\sim 35\%$) is achieved at the lowest temperature. At 200mK the threshold voltage for $I_c$ suppression moves towards larger values. 
Field effect in Al wires is observed up to $\sim 600$mK, i.e., around $40\%$ of $T_c$ (see also SI). 
The similar phenomenology obtained also in Al  FETs suggests that these effects might be intrinsic to any metallic superconducting film.  

To exclude any possible role of the substrate on the observed phenomena we have fabricated similar Ti FETs on several SiO$_2$ commercial wafers produced by different manufacturers (both doped and undoped), and sapphire substrates, without noting any substantial difference in the FETs behavior (see SI). 
Moreover, any hot spot mechanism due to direct charge injection into the wire can be ruled out as the main driving principle for $I_c$ suppression due to the incompatibility with the bipolarity of the effect, and the  independence of the critical temperature on gate voltage. In addition, a possible electron field emission mechanism can be excluded as well since it is usually expected to occur for electric fields much larger than the ones applied in the experiment.

A simplified  description of the $I_c$ behavior in our system can be obtained through a phenomenological model developed within the Ginzburg-Landau formalism \cite{Landau1950} (see SI for details). According to our model, the wire critical current can be written as 
\begin{equation}
I_c(T,V_g)=I_c^0 \left( 1-\frac{T}{T_c}\right )^{3/2}\left [1- \left (\frac{V_g}{V_g^c} \right )^4\right]^{3/2}.
\label{criticalcurrent}
\end{equation}
Figure \ref{fig2}c shows the  $I_c$ vs $V_g$ characteristics calculated from Eq. (\ref{criticalcurrent})  for the same temperatures as in panel b. 
In the calculations we used the values for $T_c$ and $V_g^c$ determined from the experiment.
Although definitely not conclusive and rather idealized, this theory is already able to grasp the essential features of our supercurrent FETs.

 %%%%%%%%%%%%%%%%%%%%%%%%%%%%%%%%%%%%%%%%%%%%%%%%%%%%%%%%%%%%%%%%%%%
\begin{figure}[t!]
\begin{center}
\includegraphics[width=\columnwidth]{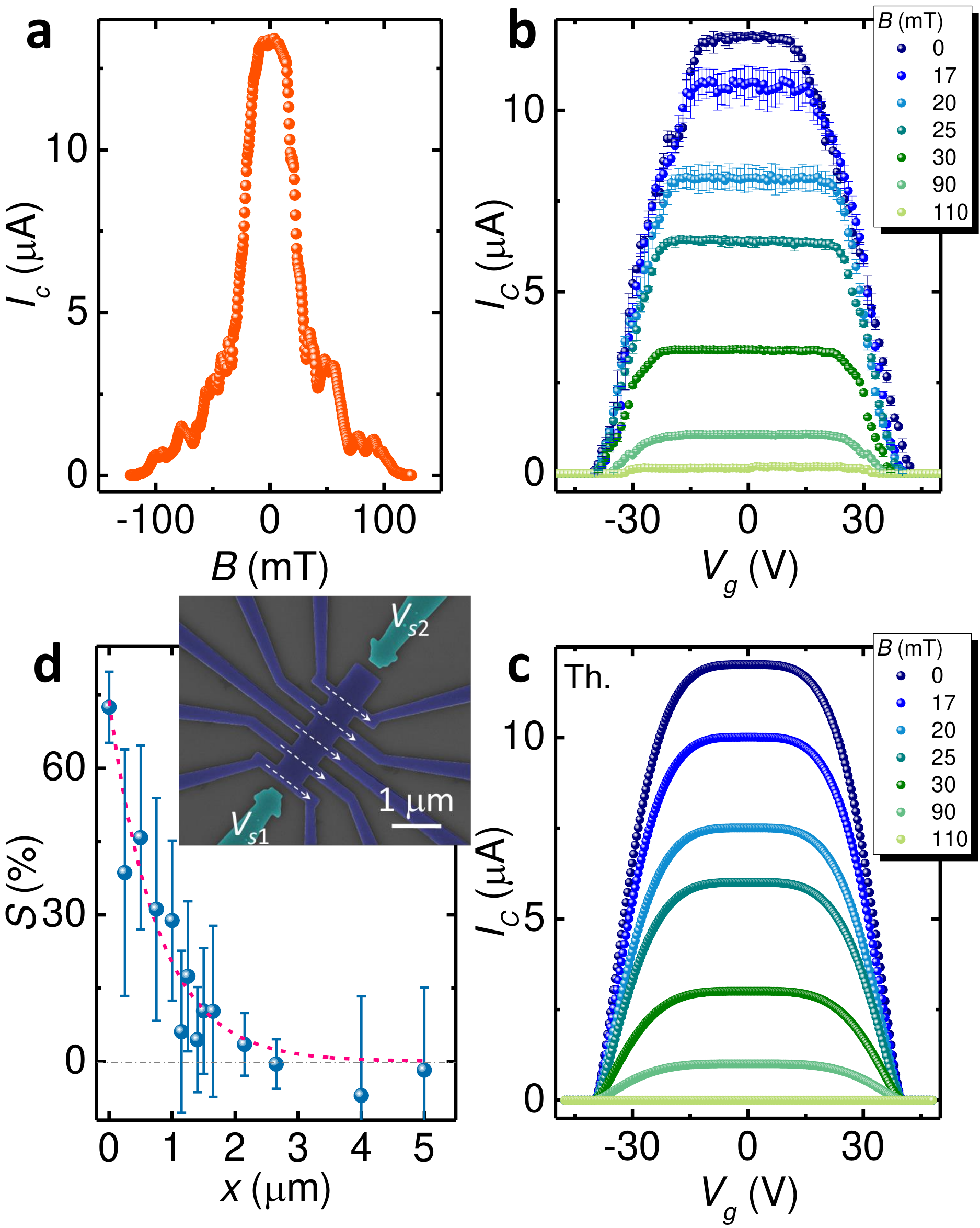}\vspace{-3mm}
\caption{\textbf{Magnetic-field dependence of the FET, and spatial extension of the electric field-induced $I_c$ suppression effect.} 
\textbf{a}, Typical pattern of the critical current $I_c$ vs perpendicular-to-plane magnetic field $B$ measured at $5\,$mK for $V_g=0$.
\textbf{b}, Behavior of $I_c$ vs $V_g$ measured at $5\,$mK for several values of the perpendicular magnetic field. 
\textbf{c},  Theoretical critical current $I_c$ vs $V_g$ characteristics calculated at $5$mK for the same values of magnetic field as in panel b.
\textbf{d}, The inset shows a pseudo-color scanning electron micrograph of a typical  Ti comb-like FET used to investigate the spatial extension of the electric field-induced $I_c$ suppression effect in the superconducting film. White dashed arrows indicate the direction of the critical current measured in each wire composing the comb.
The on-purpose asymmetry of device geometry allows to use the same wire for two $I_c$ measurements at different distances from the side gates.
The main panel displays the critical current \emph{suppression} parameter  ($\mathcal{S}$) vs distance $x$ measured at $5\,$mK for $B=0$. 
The suppression parameter represents the relative $I_c$ reduction for $V_g=90$V with respect to zero applied side gate voltage, so that $\mathcal{S}=0$ indicates the absence of field-effect induced suppression  of the critical current.
$x$ represents the average distance of the center of each wire 
 from the lateral edges of the comb. 
The critical current of each wire is recorded by polarizing one side gate at the time (i.e., $V_{s1}\neq 0$ or $V_{s2}\neq 0$) while keeping the back gate voltage fixed at $V_b=45\,$V. 
The plot summarizes all $\mathcal{S}$ values measured in two different comb-like FETs having  an average distance of  $500$nm and $1\mu$m bewteen the superconducting wires.
Dashed line is an exponential damping fit to the data [$\mathcal{S}(x)=\mathcal{S}_0 \exp (-x/\lambda)$] with a decay length $\lambda \simeq 770\pm 150\,$nm. 
The error bars in panel b and d represent the standard deviation of the critical current $I_c$ and $\mathcal{S}$, respectively, calculated over 50 measurements.
}
\label{fig3}
\end{center}
\end{figure}
%%%%%%%%%%%%%%%%%%%%%%%%%
%%%%%%%%%%%%%%%%%%%%%%%%%%%%%%%%%%%%%%%%%%
%%%%%%%%%%%%%%%%%%%%%%%%%%%%%%%%%%%%%%%%%%%%%%%%%%%%%%%%%%%%%%%%%%% 

Figure \ref{fig2}d displays the $I_c$ vs  $V_g$ characteristics measured at different  temperatures of
a square-shaped Ti transistor consisting of a 4-$\mu$m-long side, and the same film thickness. 
This structure was conceived to prove the effect also on a wide superconducting region with lateral dimensions largely exceeding $\lambda_L$ and $\xi_0$. 
The curves  reveal a similar field-effect-induced $I_c$ suppression though with reduced intensity with respect to the wires, as here only the back gate ($V_g \equiv V_b$) was applied.
Yet, this result suggests that 
the relevant spatial scale is the one parallel to the applied electric field, in analogy to the Meissner-Ochsenfeld effect. 
Therefore, even in large but thin film structures ($t\ll \lambda _L,\xi_0$), $I_c$ can be deeply affected by a perpendicular-to-plane electrostatic field.

We now discuss the joined impact of both electric and magnetic fields on the supercurrent in the wire.
The perpendicular-to-plane magnetic field dependence of the FET critical current at $5$mK and $V_g=0$ is shown in Fig. \ref{fig3}a. 
Here we note the magnetic field-induced drop of $I_c$, and its full suppression at $\sim 127$mT. 
Furthermore, the fine structure visible on the non-monotonic $I_c(B)$ characteristic might be ascribed  to the penetration of Abrikosov vortices in the Ti film upon increasing   $B$ \cite{Weber2015}. 
The full $I_c$ dependence on $V_g$ for several magnetic fields at $5$mK  is summarized in Fig. \ref{fig3}b.
All the  following characterizations were performed before any external magnetic field was applied to the transistors, unless explicitly stated.
Similarly to temperature, the effect is almost bipolar in $V_g$, and by increasing $B$ leads to a widening of the $I_c$ plateau.  
In addition, the threshold voltage $V_g^c$ turns out to be weakly dependent on $B$, in contrast to its temperature independence (see Fig. \ref{fig2}b). The reason for this different behavior is still not understood.
We note how sizable is
the effect of the electric field on  $I_c$ even in the presence of magnetic fields approaching the critical one.
Yet, our model still provides a reasonable description of the experiment, as shown in Fig. \ref{fig3}c (see SI for details).

Finally, quantification of the spatial extension of the electric field-induced non-local effect in the superconductor at $5$mK and $B=0$ is provided in Fig. \ref{fig3}d where the $I_c$  \emph{suppression} parameter ($\mathcal {S}$), defined as $\mathcal {S}=100\times [I_c(V_{si}=0)-I_c(V_{si}=90\text{V})]/I_c(V_{si}=0)$ (with $i=1,2$, see the inset of panel d displaying a representative Ti comb-like device) whilst keeping the back gate voltage fixed at $V_b=45$V, is shown. 
Differing from Figs. \ref{fig2}a,b and \ref{fig3}b, a higher $V_{si}=90\text{V}$ gate voltage value  was set to provide sizable supercurrent suppression via a single side gate.
$\mathcal {S}$ was determined for each wire composing the comb-like Ti FET designed on-purpose, and it is plotted vs the distance ($x$) between the center of each wire and the lateral
edges of the comb.  
A clear and substantial fading of field effect-induced suppression of $I_c$ is observed by increasing the distance.
Dashed line is an exponential decay fit to the data from which we extract an attenuation length $\lambda \simeq 770\pm 150$nm. 
This is intriguingly in reasonable agreement with the London penetration depth $\lambda_L\sim 900$nm previously estimated in our Ti films. 
Furthermore, measurements of $\lambda$ at different bath temperatures reveal that it is almost constant within the experimental error up to $\sim 80\%$ of $T_c$, then rapidly decreasing and vanishing by approaching the critical temperature (see SI).
The above results on the spatial extension of the field-induced non-local effect contribute to further exclude a direct heat injection into the wire at the origin of supercurrent suppression since the typical thermal relaxation length in superconductors at low temperatures is of the order of tens of microns  \cite{Fornieri2017_2}.

Our results on supercurrent FETs reveal the significant bipolar impact of a static electric field on a BCS superconducting film. 
Our physical interpretation points in the direction  of a spatial deformation of the Cooper pairing parameter driven by electric fields localized at the surface of the superconductor by conventional screening. This yields a reduction of the available net wire section able to carry a supercurrent flow.
A further step towards the understanding of the above effects might come from a set of complementary experiments, such as probing the superconductor density of states through tunneling spectroscopy, or investigating the phase rigidity by means of interference in SQUIDs, and the kinetic inductance in RF-based experiments.
These findings represent a tool to envision superconducting  field-effect devices ranging from tunable Josephson weak-links \cite{Marcus1,Marcus2}  or interferometers \cite{Clarke2004,Ronzani2014,Strambini2016} to  Coulombic \cite{Coulomb,Pekola2007} and coherent caloritronic structures \cite{Fornieri2017,Martinez2014} as well as single-photon detectors \cite{spd,Giazotto2008}, which would benefit of field effect control to enhance their functionality.  
Given the general  nature of our discovery which seems intrinsic to metallic superconducting thin films, there is flexibility for the optimization of this method both from the technological and material point of view. 
On the one hand, improved gating schemes exploiting thinner insulators could enable the reduction of gate voltage amplitudes by more than one order of magnitude. 
On the other hand, alternative superconductors such as thin NbN films ($t<10$nm) with $ T_c$ larger than $10$K,  and London penetration depth approaching $500$nm \cite{NbN1,NbN2} might be suitable candidates for the implementation of an all-metallic high-speed superconducting field-effect electronics.

\section{METHODS SUMMARY}
\subsection{Fabrication details and experimental set-up}

The supercurrent FETs were  fabricated with electron-beam lithography, and evaporation of metal through a resist mask onto a SiO$_2$/$p^{++}$-Si
commercial wafer. The deposition of Ti was performed at room temperature in an ultra-high vacuum
electron-beam evaporator with a base pressure of $\sim 10^{-10}$Torr at a deposition rate of $\simeq 11-13${\AA}/s, whereas
the deposition of Al at a deposition rate of $\simeq 1.5${\AA}/s.

The magneto-electric characterization of the FETs was performed in a filtered He$^3$-He$^4$ dry dilution refrigerator at different bath temperatures down to 5 mK using a standard 4-wire technique. The current vs voltage characteristics of the  FETs were obtained by
applying a low-noise biasing current, with voltage across the superconducting nanowire being measured
by a room-temperature battery-powered differential preamplifier. 
Gate biasing was provided by a low-noise voltage source whereas a low-frequency lock-in technique was used either for resistance vs temperature or resistance  vs magnetic field measurements.
 
The data that support the plots within this paper and other findings of this study are available from the corresponding author upon reasonable request.

\section{ACKNOWLEDGEMENTS} 
The authors wish to thank J. E. Hirsch for fruitful comments, and for having drawn the attention to relevant questions on key issues related to superconductivity so far considered well established. 
A. Braggio is acknowledged for a careful reading of the manuscript and for fruitful comments. J. S. Moodera, A. Shanenko, and P. Virtanen are acknowledged  for useful discussions.
The European Research Council under the European Union’s Seventh Framework Program (FP7/2007-2013)/ERC Grant agreement No. 615187-COMANCHE, and MIUR-FIRB2013– Project Coca (Grant No. RBFR1379UX) are acknowledged for partial financial support. 
The work of G.D.S. and F.P. was funded by
Tuscany Region under the FARFAS 2014 project SCIADRO.
The work of E.S.
was partially funded by the Marie Curie Individual Fellowship MSCAIFEF-ST No. 660532-SuperMag.
P.S. has received funding from the European Union FP7/2007-2013 under REA Grant agreement No. 630925-COHEAT.

\section{AUTHOR CONTRIBUTIONS}
G.D.S. and F.P. fabricated the samples, and performed the measurements with E. S. G.D.S. and F.P. analyzed the experimental data with input from E.S. and F.G. 
P.S. developed the theoretical model with input from F.G., and performed the numerical calculations. F.G. conceived the experiment on field effect, and wrote the manuscript with input from all authors. All authors discussed the results and their implications equally at all stages. 

\section{COMPETING FINANCIAL INTERESTS}
The authors declare no competing financial interests.

%\newpage

\section{SUPPLEMENTARY INFORMATION}

\subsection{Temperature dependence of electrostatic field attenuation length $\lambda$ in Ti FETs}

\begin{figure}[t!]
\begin{center}
\includegraphics[width=0.9\columnwidth]{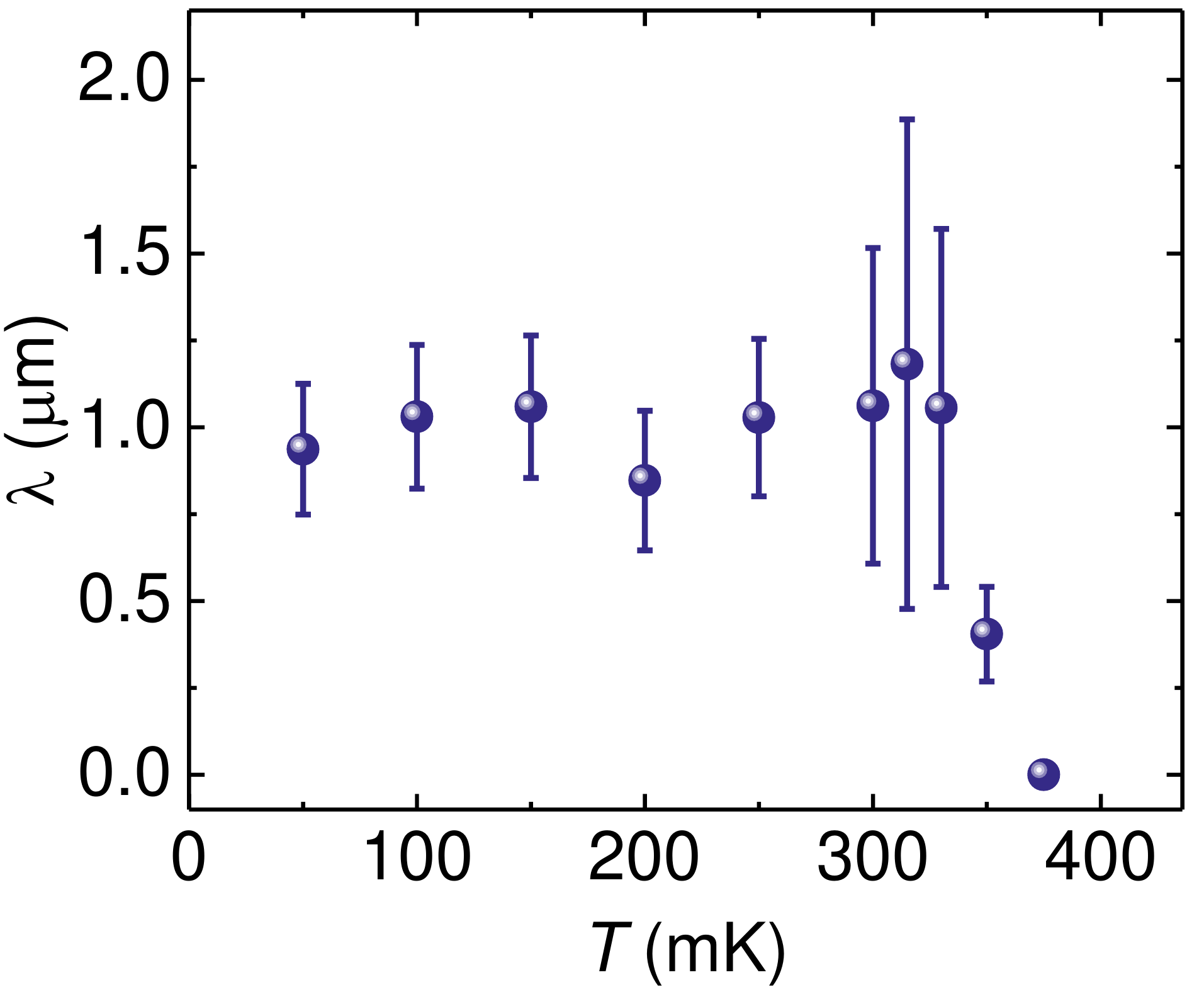}\vspace{-3mm}
\caption{\textbf{Temperature dependence of the electrostatic field attenuation length $\lambda$.} 
Behavior of $\lambda$ vs $T$ measured in a Ti comb-like FET having an average distance of 500nm between the superconducting wires. A critical temperature $T_c\simeq 430$mK characterizes the Ti film realizing this transistor. $\lambda$ is almost temperature independent up to $\sim 330$mK, and then rapidly decays to zero by approaching $T_c$.
}
\label{figS3}
\end{center}
\end{figure}

We have also measured the suppression parameter $\mathcal{S}$ as a function of distance $x$ at different bath temperatures in a comb-like Ti FET having an average distance of $500$nm between the superconducting wires. 
For each temperature, an exponential decay fit to the $\mathcal{S}(x)$ data allowed to extract the corresponding attenuation length $\lambda$.
The result of this procedure is shown in Fig. \ref{figS3} which displays the measured $\lambda (T)$ curve. 
Specifically, we note the substantial temperature \emph{independence} of the electrostatic field attenuation length  up to $\sim 330$mK, which corresponds to about  $80\%$ of the critical temperature of the Ti film composing the comb-like transistor ($T_c\sim 430$mK). A further increase of the temperature above $330$mK yields a rapid decay of $\lambda$, up to its full quenching around $375$mK. 
%%%%%%%%%%%%%%%%%%%%%%%%%%%%%%%%%%%%%%%%%%%%%%%%%%%%%%%%%%%%%%%%%%
%%%%%%%%%%%%%%%%%%%%%%%%%
%%%%%%%%%%%%%%%%%%%%%%%%%%%%%%%%%%%%%%%%%%
%%%%%%%%%%%%%%%%%%%%%%%%%%%%%%%%%%%%%%%%%%%%%%%%%%%%%%%%%%%%%%%%%%%
%%%%%%%%%%%%%%%%%%%%%%%%%%%%%%%%%%%%%%%%%%%%%%%%%%%%%%%%%%%%%%%%%%
\begin{figure}[t!]
\begin{center}
\includegraphics[width=\columnwidth]{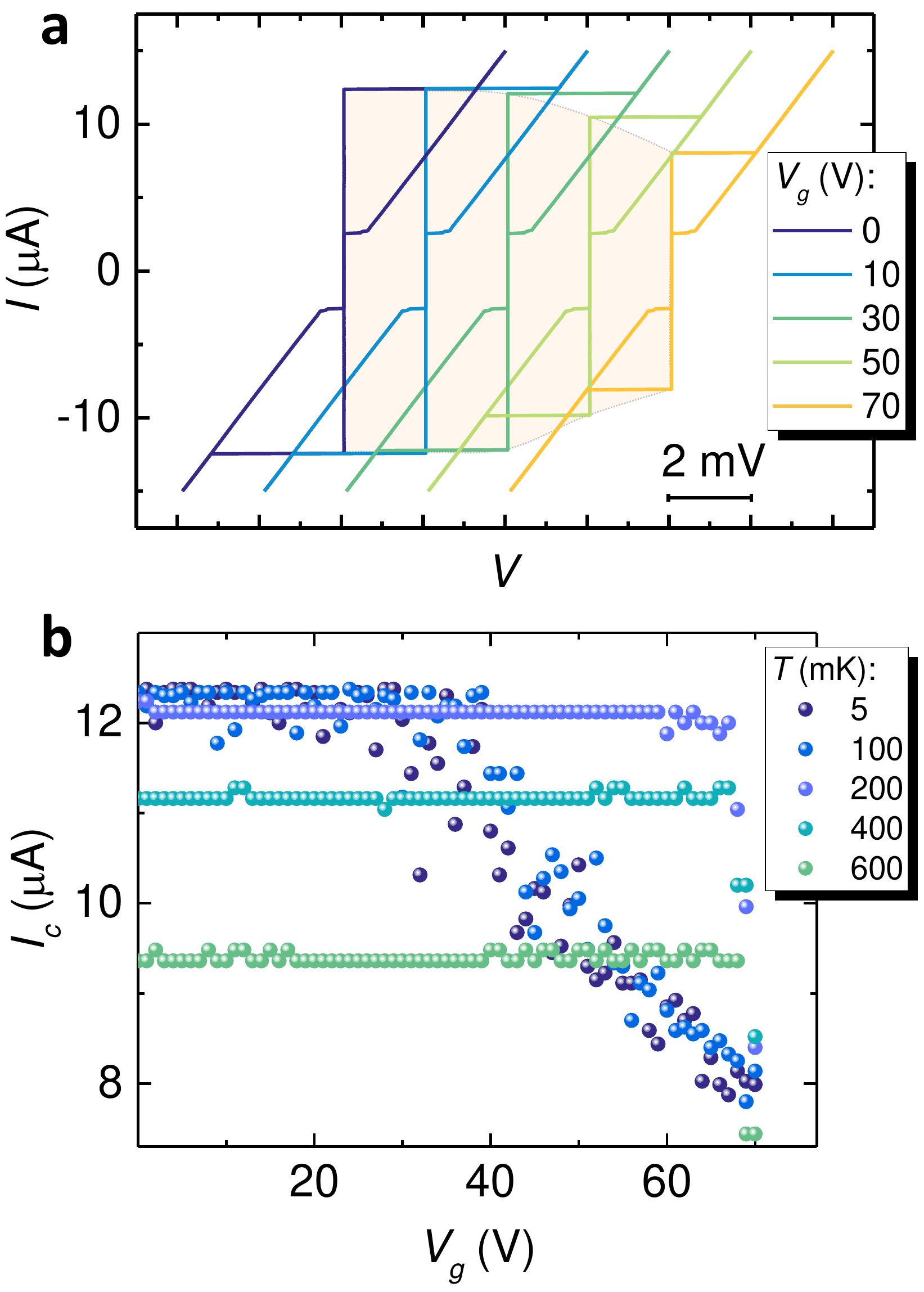}\vspace{-3mm}
\caption{\textbf{Electrostatic field dependence of the Al supercurrent FET characteristics.} 
\textbf{a}, Current-voltage ($I-V$) characteristics of an Al supercurrent FET measured at $5\,$mK for several values of the applied gate voltage ($V_g\equiv V_b$). 
Hatched area is a guide to the eye emphasizing the monotonic suppression of the critical current $I_c$ by increasing $V_g$. 
Similarly to Ti superconducting FETs, when the supercurrent vanishes the Al FET $I(V)$ characteristics  coincide with that in the normal state, independently of the applied voltage.
The curves are horizontally offset for clarity. 
\textbf{b}, Behavior of the transistor critical current $I_c$ vs $V_g$  measured at different bath temperatures $T$. 
Note the monotonic suppression of $I_c$ occurring for  $V_g\gtrsim 35$V at temperatures below $100$mK. 
This threshold voltage for supercurrent suppression largely increases at higher bath temperature.
}
\label{figS1}
\end{center}
\end{figure}

\subsection{Aluminum (Al) supercurrent FETs}

Analogous results for field effect-induced suppression of the critical current were observed in FETs made of aluminum (Al). 
In particular, the Al FETs  consist of a wire of length $l=800$nm, width $w=30$nm, and thickness $t=11$nm. 
The Al FETs were realized without side gates so that only a back gate voltage $V_b$ is used to control the supercurrent flow in the wire. 
From the normal-state resistance of the wire, $R_N\simeq 320\Omega$, and critical temperature, $T_c\simeq 1.5$K, we deduce the London penetration length $\lambda_L \simeq 310$nm,  and the superconducting coherence length $\xi_0\simeq 63$nm assuming $\mathcal{N}_F=2.15\times 10^{47}$J$^{-1}$m$^{-3}$ for the density of states at the Fermi energy of our Al films. 
For these wires holds as well the condition $w,t\ll \lambda_L$ so that substantial modulation of the critical current is expected under the application of a static electric field.

Figure \ref{figS1}a shows the wire current vs voltage ($I-V$) characteristics recorded at $5$mK of bath temperature for several values of the applied gate voltage with $V_g\equiv V_b$. 
In the present case of Al thin film transistors, the switching critical current at zero gate voltage obtains values as high as $I_c\simeq 12.3\mu$A, which is reduced down to $\sim 8\mu$A for $V_g=70$V corresponding to a relative suppression of the order of  $\sim 35\%$. 
Similarly to Ti supercurrent FETs, the current-voltage characteristics show a clear hysteretic behavior but with a larger retrapping current.
The electric field does not affect the $I-V$ characteristics when the wire switches into the normal state. 
Furthermore, differently from $I_c$, the retrapping current is unaffected by $V_g$. 
For the Al FETs it was  not possible to totally quench $I_c$ within the explored values of $V_g$.

The full temperature dependence of field-effect in the Al wire is displayed in Fig. \ref{figS1}b where the $I_c$ vs $V_g$ characteristics are shown for a few selected bath temperatures. 
In particular,  a monotonic suppression of $I_c$ occurs for $V_g \gtrsim 35$V at temperatures below $100$mK.
At higher temperature, the threshold voltage for $I_c$ suppression moves towards larger values while at the same time the critical current reduction  turns out to be somewhat limited. In the investigated range of gate voltage,  
field-effect in these Al transistors persists up to $\sim 600$mK, i.e., around $40\%$ of critical temperature. 
The reduced impact of  electric field on $I_c$ might be correlated to the smaller $\lambda _L$ (and $\xi_0$) characterizing these Al FETs in comparison to that obtained in Ti films. 

\subsection{Independence of the substrate on FETs performance, leakage current and leakage power}

\begin{figure}[t!]
\begin{center}
\includegraphics[width=\columnwidth]{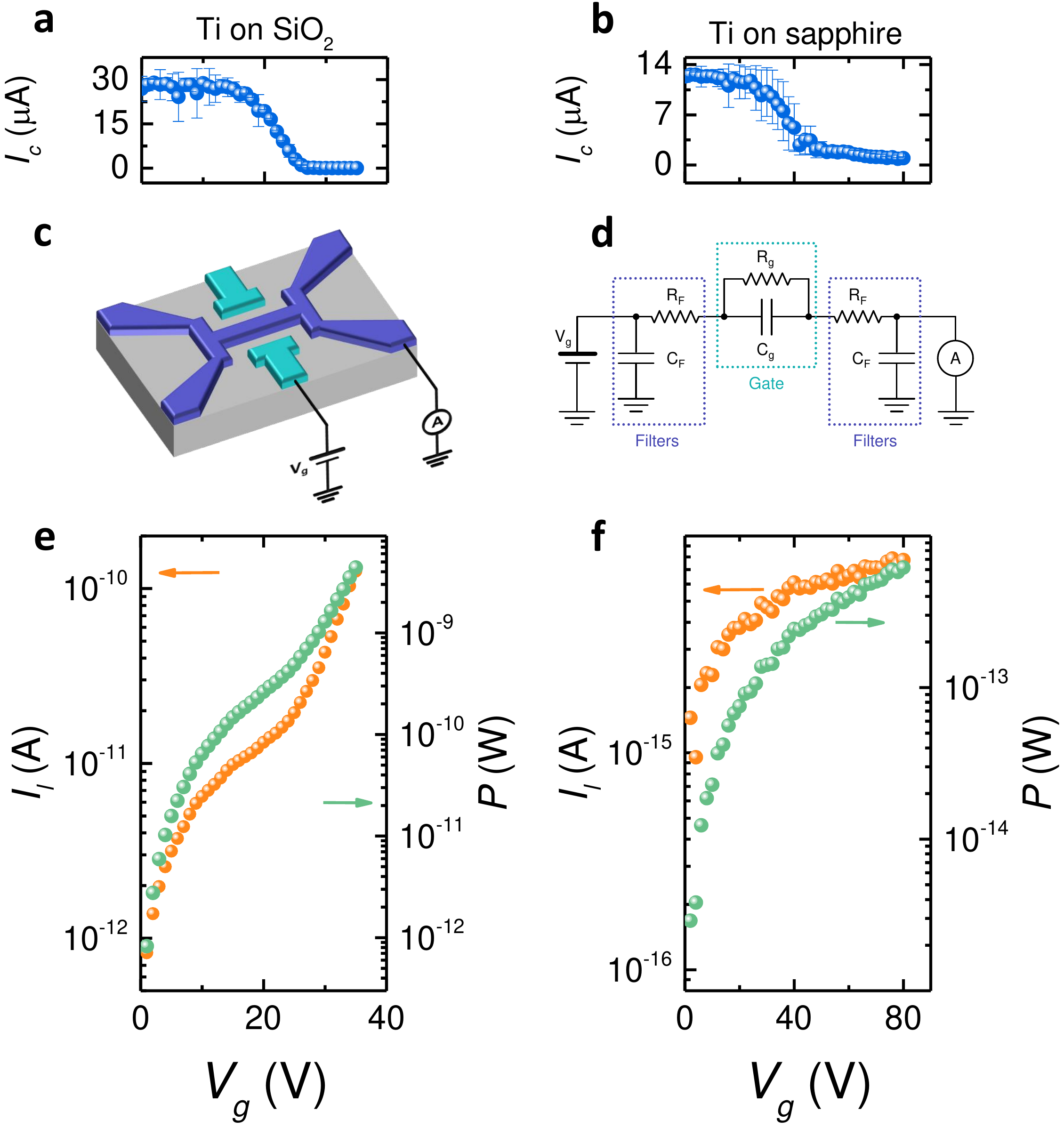}\vspace{-3mm}
\caption{\textbf{Results obtained on FETs realized on different substrates, leakage current and leakage power.} 
\textbf{a}, $I_c$ vs $V_g$ characteristic measured at 32mK of bath temperature of a similar Ti FET fabricated on a different SiO$_2$ substrate. 
\textbf{b}, The same as in panel a but for a Ti FET fabricated on a sapphire substrate. The two devices show somewhat similar behavior which resembles that one of the Ti FET described in the main text of the paper. 
\textbf{c}, Scheme of the measurement setup used to characterize the leakage currents $I_l$.
\textbf{d}, Complete equivalent electrical circuit of the setup used for the measurement of $I_l$ vs $V_g$. $C_F$ and $R_F$ are the capacitor and the resistor used for the RF filtering of the cryostat, respectively. $C_g$ and $R_g$ are the effective capacitance and resistance between the gates and the wire, respectively. 
\textbf{e}, Leakage current $I_l$ vs $V_g$ (left vertical axis, orange dots), and corresponding leakage power $P$ vs $V_g$ (right vertical axis, green dots) for the same device of  panel a.
\textbf{f} The same as in panel e but for the FET fabricated on sapphire substrate shown in panel b. The error bars in panel a and b represent
the standard deviation of the critical current $I_c$
calculated over 50 measurements.
}
\label{figS4}
\end{center}
\end{figure}

To exclude that the type of substrate has some role in the origin of the observed field effect we have also fabricated Ti supercurrent transistors on Si wafers  produced by different manufacturers (both doped and undoped), and on sapphire substrates. 
Figure \ref{figS4} displays in the  top panels the $I_c$ vs $V_g$ characteristics recorded at $32$mK of bath temperature of a similar Ti FET deposited onto a SiO$_2$ wafer (panel a), and onto a sapphire substrate (panel b). Both FETs lack the backgate so that only side gates were used to control the supercurrent.
In particular, besides the difference in the maximum critical current at zero gate bias, both FETs show the usual monotonic suppression of $I_c$ by increasing $V_g$. 
The slightly different characteristics as a function of $V_g$ stems from differences existing in the exact geometry of the two transistors. 
The comparison of the behavior of the Ti FETs shown in panels a and b of Fig. \ref{figS4},  and the one displayed in the main text of the paper confirms that the observed field effect is completely  \emph{independent} of the type of substrate used to realize the supercurrent transistors.

%%%%%%%%%%%%%%%%%%%%%%%%%%%%%%%%%%%%%%%%%%%%%%%%%%%%%%%%%%%%%%%%%%

In addition, in characterizing the FETs we have measured the existing leakage current ($I_l$) present in the whole measurement circuit upon biasing the transistor gates, in order to assess the corresponding leakage power. 
Left vertical axes of Fig. \ref{figS4} c and d show the $I_l$ vs $V_g$ characteristics (orange dots) of the transistors of panel a and b, respectively. 
Specifically, $I_l$ obtains values as large as a few $\sim 10^{-11}$A in the region of supercurrent suppression for FETs realized on SiO$_2$, whereas of the order of several
 $\sim 10^{-15}$A for supercurrent transistors realized on sapphire substrates. 
The above values are standard leakage currents  measured in all the devices we realized on different substrates. 
From this follows that in the former case $I_l$ is typically around $\sim 10^{-6}I_c(V_g=0)$ while for the latter the typical leakage current is around $\sim 10^{-10}I_c(V_g=0)$. 

The corresponding  leakage power in the whole circuit due to losses, i.e.,  $P=I_lV_g$, is displayed in panels c and d for both transistors  as green dots (right vertical axes). 
In particular, the typical maximum leakage power $P$ for FETs realized onto SiO$_2$ is of the order of a few  $\sim 10^{-10}$W at complete supercurrent suppression (panel c) whereas it obtains values as large as a few $10^{-13}$W in the same regime for FETs fabricated on sapphire (panel d).

We finally wish  to emphasize that the above leakage power cannot be considered at the origin of supercurrent suppression in our FETs stemming from quasiparticle overheating due to direct injection of electric  current.  
As a matter of fact,  it can be shown through numerical calculations of the thermal steady-state of a superconductor under power injection that any quasiparticle overheating leading to a reduction or complete suppression of $I_c$ would reflect into a sizable decrease of the wire critical temperature $T_c$. 
Measurements of critical temperature  performed on different wires under $V_g
$  conditions leading to full suppression of $I_c$ have shown that $T_c$ is completely \emph{independent} of gate voltage  within the experimental error (see data in \cite{Paolucci2018}), thereby supporting the above claim.

\subsection{Phenomenological theory of field effect-induced suppression of the critical supercurrent $I_c$}

In order to calculate the impact of the electric field on the critical current, we start from the
 the superconducting free energy density $F$, which in the absence of magnetic field, reads \cite{Landau1950_SI, degennes_SI, Tinkham1_SI}:
\begin{equation}
F = F_n + \alpha(T) |\psi|^2 + \frac{\beta}{2} |\psi|^4 + \frac{\hbar^2}{2 m } \Big|  \nabla  \psi \Big|^2 + \Lambda \frac{\mathE_{tot}^2}{8 \pi}.
 \label{eq:GL_free_energy_Ef}
\end{equation}
Following Ginzburg-Landau theory, $\alpha = \alpha_0 (T/T_c -1)$ and $\beta$ parameters describe the transition from superconducting to normal state.
The superconducting state is obtained for $\alpha<0$ and order parameter minimum is found for $\psi = \psi_0 = -\alpha/\beta$ \cite{degennes_SI, Tinkham1_SI}.
The term $F_n$ represents the free energy density in the normal state, whereas the last term is the energy density associated to the total electric field $\mathE_{tot}$.
The parameter $\Lambda$ takes into account the penetration of the electric field into the superconductor; it can vary between $\Lambda = 0$ for no penetration and 
$\Lambda =1$ for a fully penetrated superconductor.
The value of $\Lambda$ is presently unknown, and is supposed to be almost independent of $\psi$ and $\nabla \psi$.
From the experimental measurements we know that, while the critical current is strongly affected by the electric field, the wire critical temperature remains constant (see data in \cite{Paolucci2018}). This poses a strong constraint on the phenomenological modelization within the Ginzburg-Landau theory. The electric field cannot affect neither $|\psi|^2$ nor $|\psi|^4$ term in the free energy (\ref{eq:GL_free_energy_Ef}) since this would result in a change of  the critical temperature \cite{degennes_SI, Tinkham1_SI}. Therefore, the only way the electric field can affect the condensate is through the $\nabla  \psi$ that describes the spatial deformation of the order parameter. The microscopic reason for such a condensate deformation is, at present, unknown. 
It could be due to the non-local propagation of the perturbation induced by  surface electric fields or to the formation of inhomogeneous regions inside the superconductor. Below we show that such microscopic derivation is not essential within the framework of the Ginzburg-Landau theory, providing a simple phenomenological model able to catch the main features of the observed physical phenomenon and assuming a field dependent deformation of the order parameter.

\begin{figure}
    \begin{center}
    \includegraphics[width=\columnwidth]{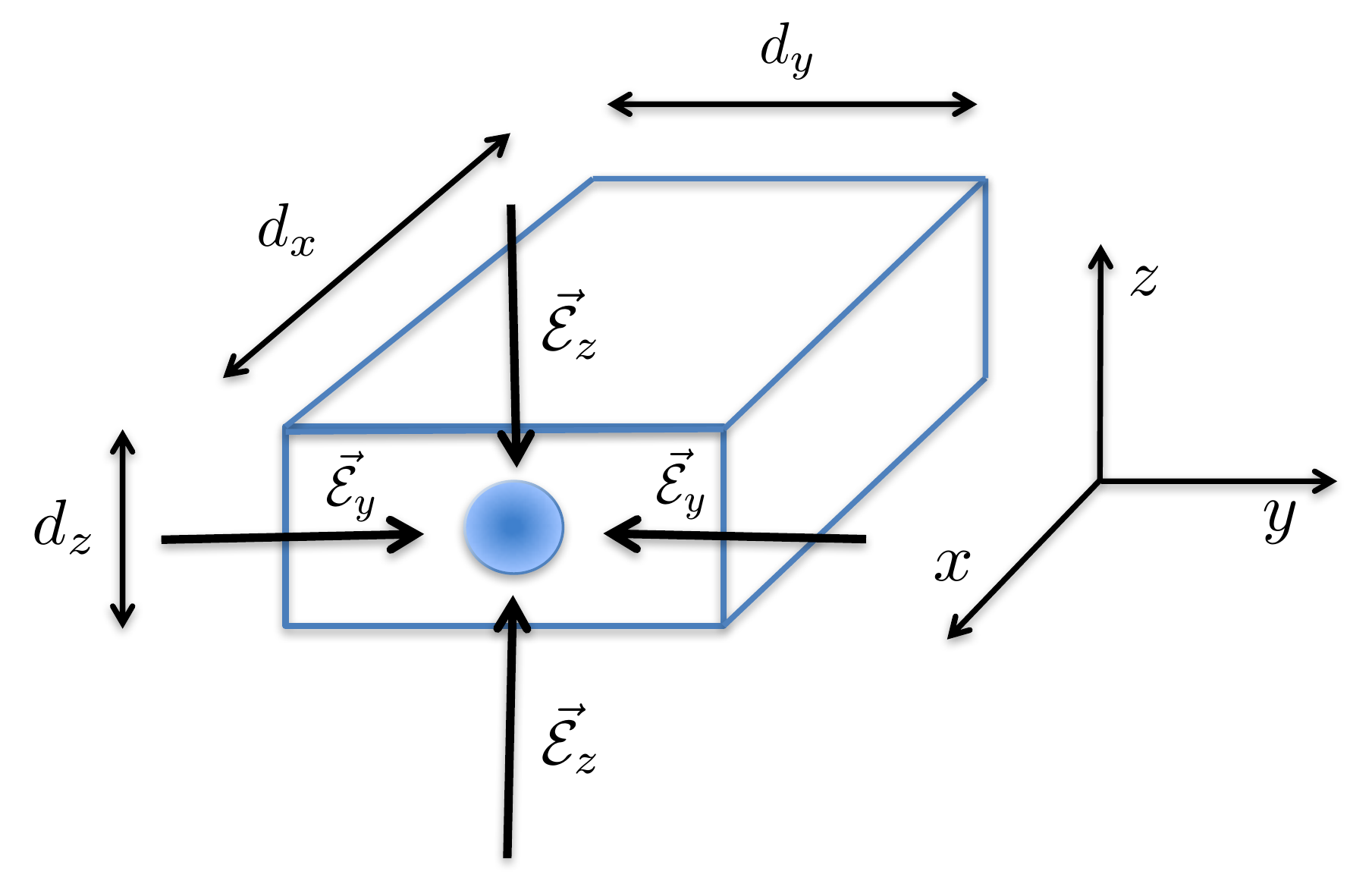}\vspace{-3mm}
       \end{center}
    \caption{{\bf Effect of electric fields on a superconducting wire.}
     A superconducting wire of length $d_x$, $d_y$ and $d_z$ is subject to electric fields $\vec{\mathE}_y$ and $\vec{\mathE}_z$. The electric fields produce a deformation of the Ginzburg-Landau order parameter (shown as a blue circle) along the field directions. 
    The current flows along the $x$ direction.  
    }  
    \label{wire}
\end{figure}

We consider a wire with length $d_x$ and lateral dimensions $d_y$ and $d_z$ (with $d_x \gg d_y, d_z$) and, accordingly, we set the coordinate axis (see Fig. \ref{wire}).
The current flows along the $x$ direction and the extremes of the wire are at $y=\pm d_y/2$ and $z=\pm d_z/2$.
The electric fields are applied along the $y$ and $z$ directions.

We write the order parameters as $\psi = \psi_x \psi_y \psi_z$.
Along the current flowing direction $\psi$ is uniform, i.e., $|\psi_x| = \psi_0 f$ where $f$ is a space-independent parameter keeping into account the modulation of $\psi$.
Our working assumption is that the presence of the electric field induces \emph{deformation} the order parameter $\psi$ along  $y$ and $z$ directions.
This deformation of $\psi$ can be due either to a direct influence of the electric field inside the superconductor, or to the propagation of the electric perturbation occurring at the surface of the wire.
To simplify the treatment, this deformation is described by Gaussian functions centered in $y=0$ and $z=0$, and with standard deviation $\sigma_y$ and $\sigma_z$.
Accordingly,  we have
\begin{equation}
 \psi  = \psi_0 f ~    \Big(\frac{e^{- \frac{ y^2}{2 \sigma_y^2}}}{\sqrt{2 \pi} ~\sigma_y/d_y} \Big)^{\frac{1}{2}}  ~    \Big( \frac{e^{- \frac{ z^2}{2 \sigma_z^2}}}{\sqrt{2 \pi} ~\sigma_z/d_z}\Big)^{\frac{1}{2}} .
 \label{eq:GL-wave-function}
\end{equation}
An increase of the electric field $\mathE_i$ along the $i$-th direction leads to a deformation of the order parameter along the same direction and to a decrease of $\sigma_i$.
To describe this effect we assume that  $\sigma_i = \mathcal{V}/\mathE_i$ where $\mathcal{V}$ has the dimension of a voltage.
%%%%%%%%%%%%%%%%%%%%%%%%%%%%%%%%%%%%%%%%%%%%%%%%%%%%%%%%%%%%%%%%%%%%%%%%%%%%%%%%%%%
%%%%%%%%%%%%%%%%%%%%%%%%%%%%%%%%%%%%%%%%%%%%%%%%%%%%%%%%%%%%%%%%%%%%%%%%%%%%%%%%%%%
%%%%%%%%%%%%%%%%%%%%%%%%%%%%%%%%%%%%%%%%%%%%%%%%%%%%%%%%%%%%%%%%%%%%%%%%%%%%%%%%%%%%%%

Keeping track of the spatial modulation of $\psi$, the free energy density (\ref{eq:GL_free_energy_Ef}) becomes
\begin{equation}
F = F_n + \alpha \psi^2 + \frac{\beta}{2} \psi^4 + \frac{\hbar^2}{8 m } \left(\frac{y^2}{ \sigma_y^4} + \frac{z^2}{ \sigma_z^4}\right) \psi^2 + \Lambda \frac{\mathE_{tot}^2}{8 \pi} .
\end{equation}
By minimizing the latter with respect to $\psi$, the electric field contribution vanishes and we arrive at the equation $ \alpha + \beta \psi^2 + \frac{\hbar^2}{8 m } \left(\frac{y^2}{ \sigma_y^4} + \frac{z^2}{ \sigma_z^4}\right)=0$. Using the relations $\alpha = - |\alpha|$ and  $\psi_0 = |\alpha|/\beta$, we obtain
\begin{equation}
 -|\alpha| (1-  f^2 \psi_y^2 \psi_z^2) = \frac{\hbar^2}{8 m } \left(\frac{y^2}{ \sigma_y^4} + \frac{z^2}{ \sigma_z^4}\right).
\end{equation}
We take the average over $y$ and $z$ directions, i.e., $1/d_y \int_{-d_y/2}^{d_y/2} dy$ and $1/d_z \int_{-d_z/2}^{d_z/2} dz$ .
When the order parameter is strongly suppressed, i.e., 
for $\sigma_i \ll d_i$, we have that $1/d_y \int_{-d_y/2}^{d_y/2}  \psi_y^2 dy\approx 1/d_y \int_{-\infty}^{\infty}  \psi_y^2 dy=1$ and $1/d_y \int_{-d_y/2}^{d_y/2}  y^2 dy =d_y^2/12$ and analogously for the $z$ coordinate. 

For a symmetric wire, and equal applied electric fields we have  $d_y=d_z=d$, $\mathE_y=\mathE_z=\mathE$, $\sigma_y=\sigma_z=\sigma$ and $f^2 = 1- \frac{\hbar^2}{48 m |\alpha|} \frac{d^2}{ \sigma^4}$ that can be written as 
\begin{equation}
f^2 =  1- \frac{1}{\bar{\alpha}(T)}\Big( \frac{\mathE}{\mathE_c}\Big)^4 
\label{eq:f_critical_field}
\end{equation}
where we have supposed that $\sigma = \mathcal{V}/\mathE$, and defined $\bar{\alpha}(T) = |\alpha (T)|/\alpha_0 = (T_c-T)/T_c$ and $\alpha_0 \equiv \alpha(T=0)$ that has the dimension of an energy.

In equation (\ref{eq:f_critical_field}) we have introduced an electric critical field $\mathE_c = [48 m \alpha_0 V^4/(\hbar^2 d^2)]^{1/4}$.
Since $0 \leq \psi \leq \psi_0$, we have that $0 \leq f \leq 1$. Equation (\ref{eq:f_critical_field}) tells us that if $\mathE \geq \bar{\alpha}^{1/4} \mathE_c $, the superconducting state is destroyed since there is no $f$ that can satisfy it
\cite{degennes_SI, schmidt1997physics_SI}.

The electric field $\mathE$ in Eq. (\ref{eq:f_critical_field}) is the \emph{effective} field affecting the superconductor and generating the order parameter deformation.
This could be different from the externally applied electric field $\mathE_{ext}$; for this reason we take $\mathE = \chi \mathE_{ext}$.

It is natural to assume that $\chi$ depends on temperature.
At $T=0$ we assume that $\chi(T=0)=1$ and $\mathE = \mathE_{ext}$.
At the superconducting critical temperature, the system becomes normal, and the external electric field has no effect on the metal. Therefore, we have $\chi(T=T_c)=0$ and $\mathE = 0$.
With these observations, we assume a temperature dependence of $\chi$ as follows:
\begin{equation}
 \chi(T) = \left( 1- \frac{T}{T_c} \right)^\eta = \bar{\alpha}^\eta,
 \label{eq:chi}
\end{equation}
and we obtain a modified equation for $f$
\begin{equation}
f^2 =  1- \frac{1}{\bar{\alpha}^{1-4 \eta}}\Big( \frac{\mathE_{ext}}{\mathE_c}\Big)^4. 
\label{eq:f_critical_field_screening}
\end{equation}

From Eq. (\ref{eq:f_critical_field_screening}), we can infer a phenomenological behavior of the critical current.
In the absence of magnetic and electric fields, the critical current reads  $I_c = \frac{8}{3 \sqrt{3}} \frac{e}{\sqrt{m} \beta} |\alpha|^{3/2}$ \cite{degennes_SI, Tinkham1_SI}. 
The information about the suppression of superconductivity can be included by modifying the critical current [normalized to $I_c^0 \equiv I_c(T=0, \mathE= 0)$] as
\begin{eqnarray}
 \frac{ I_c}{I_c^0 } &=&  \bar{\alpha}^{3/2} f^{2\gamma}
 \nonumber \\
 &=& \Big(1-\frac{T}{T_c}\Big)^{3/2} \Big [ 1- \frac{1}{(1-T/T_c)^{1- 4 \eta}} \Big( \frac{\mathE_{ext}}{\mathE_c}\Big)^4 \Big ]^\gamma
  \label{eq:phenomenological_jc}
\end{eqnarray}
where $\eta$ and $\gamma$ are parameters to be determined from the experiment. 
We can rewrite the critical current expression as a function of a critical voltage ($V_g^c$) with $\mathE_{ext}/\mathE_c  = V_g/V_g^c$.
The plots in Fig. $2$c of the main text are obtained for $\eta=1/4$ and $\gamma = 3/2$.

\subsection{Critical current with electric and magnetic field}

The description of the critical current behaviour in presence of both electric $\mathcal{E}$ and magnetic field $B$ is more complex.
It is known  that the presence of a magnetic field reduces the critical current and the standard theory predicts a decrease with $B$ (see Refs. \cite{Bardeen1962_SI,Mydosh1965_SI, degennes_SI, schmidt1997physics_SI, Tinkham1_SI}).

However, as shown in Fig. $3$a of the main text, the experimental measured values of the critical current at $\mathE=0$ have a complex dependence on $B$.
In particular the oscillator behavior at high magnetic field, are related to the Weber blockade \cite{Morgan-Wall2015_SI} that is not accounted in the standard theory \cite{Bardeen1962_SI, Mydosh1965_SI, degennes_SI, schmidt1997physics_SI}.
For this reason, here we adopt a more practical approach.

From the previous discussion we know that the critical current is suppressed as $(\mathE_{ext}/\mathE_c)^4$.
To include the magnetic field dependence, we use the phenomenological function
\begin{equation}
 \frac{ I_c}{I_c^0 } = I_{meas}(B) \Big [ 1-  \Big( \frac{\mathE_{ext}}{\mathE_c}\Big)^4 \Big ]^\gamma =I(B) \Big [ 1-  \Big( \frac{V_g}{V_g^c}\Big)^4 \Big ]^\gamma
 \label{eq:phenomenological_jc_both_fields}
\end{equation}
where $ I_{meas}(B) \equiv I_{meas}(B, \mathE=0)$ is the (normalized) value of the critical current measured at $B$ and $\mathE=0$.\
Since the measurements are done at constant and low temperature ($T=10~$mK corresponding to $T/T_c = 0.03$), we can assume $\bar{\alpha} \approx 1$.
The plots in Fig. $3$d of the main text are obtained from Eq. (\ref{eq:phenomenological_jc_both_fields}) with $\gamma=3/2$.

%%%%%%%%%%%%%%%%%%%%%%%%%%%%%%%%%%%%%%%
\subsection{Supplementary Information  References}

%%%%%%%%%%%%%%%%%%%%%%%%%%%%%%%%%%%%%%%
 
\end{document}